\documentclass[11pt,twoside]{article}
\usepackage{epsfig}

\newcommand{\lab}[1]{\label{#1}}
\newcommand{\re}[1]{(\ref{#1})}
\newcommand{\bfr}{\begin{flushright}}
\newcommand{\bfl}{\begin{flushleft}}
\newcommand{\efl}{\end{flushleft}}
\newcommand{\efr}{\end{flushright}}
\newcommand{\bc}{\begin{center}}
\newcommand{\ec}{\end{center}}
\newcommand{\be}{\begin{equation}}
\newcommand{\ee}{\end{equation}}
\newcommand{\bea}{\begin{eqnarray}}
\newcommand{\eea}{\end{eqnarray}}
\newcommand{\ba}{\begin{array}}
\newcommand{\ea}{\end{array}}
\newcommand{\edc}{\end{document}}

\newcommand{\ds}{\displaystyle}
\newcommand{\dsf}{\displaystyle\frac}

\begin{document}

\title{Implicit Anyon or Single Particle Boson Mechanism of HTCS and Pseudogap Regime}

\author{B. Abdullaev  \\
Institute of Applied Physics, National University of Uzbekistan,\\
 Tashkent 700174, Uzbekistan (e-mail: babdullaev@nuuz.uzsci.net)\\
and\\
Research Center for Dielectric and Advanced\\
Matter Physics,Pusan National University, Gumjung, Pusan 609-735,\\
Republic of Korea (e-mail: abdullaev@pusan.ac.kr)}


\gdef\theauthor{B. Abdullaev }
\gdef\thetitle{Implicit anyon or single particle boson
...}

\date{}

\maketitle

\begin{abstract}
We propose a single particle boson mechanism of High $T_c$ Superconductivity (HTCS) and pseudogap regime.
Bosons appear in it due to the coupling  of spins of the two-dimensional (2$D$) fermions with statistical magnetic field
induced by anyon vector potential. The ground state of 2$D$ gas is pure bosonic if gas is not dense.
At the dense limit of gas the interaction of effective (coupled with the statistical magnetic field)
spins of bosons leads to the increasing of their fluctuations, which destroy the coupling. An
experimental phase diagram of the hole doped  superconducting cuprates discussed in the paper of Tallon
and Loram might qualitatively and quantitatively be clarified in the framework of this mechanism. The vicinity
of the structural phase transition to superconducting  state might strengthen the possible quadratic striction
in the sample and the phase transition of bosons into Bose-Einstein condensate (BEC), which is responsible for
the superconductivity (SC), is not second order, but first, close to second one.  According this treatment the
pseudogap regime is the region of meta stable bosons, which are out of the BEC. At the pseudogap boundary, $E_g$,
the bosons finally undergo the phase transition into fermions. Non-Fermi liquid like property of quasi-particles
discussed in the literature might be related  to bosons with spins in the pseudogap regime.
\end{abstract}

\section{Introduction}

A mechanism of SC based on the BEC of bosons, irrespective of nature of particles, has been proposed by Ogg,
Ref.  \cite{ogg},in 1946 and then  by Schafroth together with  collaborators, Refs. \cite{shaf,shafbb},
(see also the review of Ginzburg, Ref. \cite{ginzburg} ) during the decade since 1950.
It seems a single particle boson nature of SC is mysterious, because no reason for these
particles to appear in the solid materials. Only astrophysical objects, like stars, allow \cite{ginzburg}
at the extremal physical conditions this scenario of SC. It is believed \cite{dagotto} that, as metals at
low temperatures, the HTCS materials are described by Cooper like pairs of fermions and the set of 2$D$
superconducting planes of atoms $CuO_2$ plays the important role. For these 2$D$ systems has been suggested the
alternative semion (anyons with fractional parameter $\nu=1/2$) picture of HTCS, which was latter not
supported in the experiment ( the references on the theoretical and experimental papers with respect to the
subject see in Ref. \cite{ler}). The extended discussion of this treatment, where anyons are the quasi-particle
excitations, is outlined in \cite{wil1}.

In general, the concept of anyons, as result of non-relativistic Chern-Simons Quantum Field theory \cite{ler},
discovers the physical richness of two space dimensions. The topology of configuration space for orbital motion
of particles in 2$D$ allows  for fractional exchange statistics  \cite{lei}, characterized by a continuous parameter
$\nu$ that may attain values between 0 (for bosons) and 1 (for fermions). Particles with $0<\nu<1$ are generically
called anyons \cite{wil2}.

Assuming anyons are spinless, it is believed \cite{ler} that particles in this concept
can continuously  transform from canonical bosons  into canonical fermions \cite{ortiz}. This means, for instance,
the symmetric wave function of boson ground state can continuously transform into antisymmetric one of fermion
ground state. In the paper we follow this notion.

The interesting and not yet considered in the literature on anyons problem is the  relation of fractional
statistics and real spins of particles. From standard courses of Quantum Electrodynamics (see, for example,
Ref. \cite{berestecky}) it is well known that particles with the integer number of $\hbar /2$  spins possess a
Fermi statistics with Pauli exclusion principle for occupation of one quantum state by particle and antisymmetry
constraint for the many-body wave function. In the same time, for 2$D$ systems the  concept of anyons (at $\nu=1$)
gives an opportunity to introduce this antisymmetry  property of the  wave function into Hamiltonian. In the paper
we assume that anyons  (like electrons or holes) have the spin $\hbar /2$.

The goal of the  present paper is to outline the results of paper \cite{abdullaev}, where we have studied the
simultaneous effect of spins and fractional statistics and found that the value $\hbar /2$ is crucial for 2$D$ systems.
We have introduced in the Hamiltonian of anyon gas the Zeeman term of the  interaction of spins of particles
with magnetic field induced by anyon vector potential, i.e. the statistical magnetic field \cite{ler,dunne}, and showed
that the calculation of an expectation value for ground state energy exhibits the total cancellation  of terms
connected with statistics. As the cancellation occurs at any $\nu \not=0$ this would mean the bosonization of anyons
and, at particular case $\nu=1$, of 2$D$ fermions in the ground state. Expecting that this effect would be general for
any 2$D$ gas of fermions, we have applied it for the interpretation of phase diagram of HTSC and pseudogap regime
suggested by Tallon and Loram in the review of experimental papers,  Ref. \cite{tallon}.

Previously, we have derived an approximate analytic expression for the ground
state energy of $N$  charged anyons confined in a 2$D$ harmonic potential
\cite{aormn}. This was achieved by using the bosonic representation of anyons and
a gauge vector potential to account for the fractional statistics, which allowed
working with the product ansatz for the $N$ -body wave function. A variational
principle  has been applied  by constructing this wave function from
single-particle gaussians of variable shape. As in many other perturbative
treatments of anyons in an oscillator potential (see references
in Ref. \cite{aormn}) our expression for the ground state energy had a logarithmic
divergence connected with a cut-off parameter for the interparticle distance.
Making use of the physical argument (see Ref.~\cite{ler}) that for $\nu\neq 0$
this distance has to have some finite value, we have regularized the formula
obtained for the ground state energy by an appropriate procedure that takes into
account the numerical results for electrons in quantum dots in the case with
Coulomb interaction.

In our treatment, Ref. \cite{abdullaev},  the normal and superconducting states were separated
by gap -- the difference of the ground state energies of fermions and bosons and thus, at the evaluation
of phase diagram for BEC we have used the approximate analytic expressions for these energies
obtained in our recent paper \cite{aom} for the  2$D$ homogeneous Coulomb Fermi and Bose gases.

The paper is organized as follows. After brief discussion in the Section 2 of our approach and results
\cite{aormn} for the confined in harmonic potential anyons with Coulomb interaction, in the Section 3 we describe
the harmonic potential regularization procedure \cite{aom} to get the thermodynamic limit and then obtain the
approximate analytic expression for the ground state energy of the 2$D$ homogeneous Coulomb anyon gas.
The results of paper \cite{abdullaev} about the implicit anyon or sigle particle boson  mechanism of HTCS and
pseudogap regime will be outlined in the Section 4 and we summarize and conclude the present paper by the
Section 5.

\section{The Coulomb interacting anyons in a $2D$ harmonic potential}

The Hamiltonian of $N$ spinless anyons of mass $M$ and charge $e$
confined to a $2D$ harmonic potential, interacting through Coulomb
repulsions, is given by
\be
\hat H=\dsf{1}{2M}\ds\sum_{k=1}^N\left[\left(\vec p_k+\vec
A_{\nu}(\vec r_k)\right)^2+M^2\omega_0^2
|\vec{r_k}|^2\right] +\dsf{1}{2}\ds\sum_{k,j\not=k}^N\dsf{e^2}{|\vec
r_{kj}|} \ .
\lab{gsetup1}
\ee
Here $\vec r_k$ and $\vec p_k$ represent the position and momentum
operators of the $k$th anyon in two space dimensions,
\be
\vec A_{\nu }(\vec r_k)=\hbar\nu\ds\sum_{j\not=k}^N\dsf{\vec e_z
\times\vec r_{kj}} {|\vec r_{kj}|^2}
\lab{gsetup2}
\ee
is the anyon gauge vector potential \cite{wu,lau}, $\vec r_{kj}=\vec
r_k-\vec r_j$, and $\vec e_z$ is the unit vector normal to the 2$D$
plane. The factor $\nu$ determines the fractional statistics
of the anyon: it varies between $\nu=0$ (bosons) and $\nu=1$
(fermions).

We employ a variational
scheme by minimizing the expression for the total energy
\be
E=\dsf{\int \Psi_T^*(\vec R)\hat H \Psi_T(\vec R) \ d\vec R}{\int
\Psi_T^*(\vec R) \Psi_T(\vec R) \ d\vec R}
\lab{gsetup3}
\ee
with a trial wave function $\Psi_T(\vec R)$ depending on the
configuration $\vec R = \{\vec r_1....\vec r_N\}$ of the $N$ anyons.

It is reasonable, in the bosonic representation of anyons
when the many-body wave function takes the product form
\be
\Psi_T (\vec R)=\prod_{k=1}^N\psi_T (\vec r_k)
\lab{gsetup5}
\ee
to adopt the single-particle trial functions $\psi_T(\vec r_k)$ in the
form
\be
\psi_T(\vec
r_k)=C\exp\left(-(\alpha'+\nu)\dsf{(x_k^2+y_k^2)}{2L^2}\right) \ .
\lab{gsetup6}
\ee
Here $C$ is a normalization
constant and $\alpha'$ a variational parameter. We identify $L$ with the
characteristic length $(\hbar/M\omega_0)^{1/2}$ of the harmonic
oscillator.

When energies are expressed in units of $\hbar\omega_0$ and
lengths in units of $L$ the normalized trial wave function reads
\be
\Psi_T(\vec R)=\left(\dsf{\alpha}{\pi}\right)^{N/2}\prod_{k=1}^N
\exp\left(-\alpha\dsf{(x_k^2+y_k^2)}{2}\right) \ ,
\lab{gsetup7}
\ee
where $\alpha=\alpha'+\nu$.

In evaluating the expectation value $E$ (Eq.~\re{gsetup3}) it is
convenient to consider the local energy  $E_L(\vec R)=\Psi_T^{-1}(\vec
R)\hat H \Psi_T(\vec R)$ \cite{ceperley}. In general $E_L(\vec R)$ is a
complex function
\be
E_L(\vec R)=Re E_L(\vec R)+i Im E_L(\vec R)
\lab{gsetup8}
\ee
with
\be
Im E_L(\vec R) = - \alpha\ds\sum_{k=1}^N((\vec A_{\nu}(\vec
r_k)+ e\vec{A}_{ext}(\vec r_k)/c)\cdot\vec r_k) \ .
\lab{gsetup9}
\ee
However, evaluation of the expectation value $E=\int \Psi_T(\vec R)\
E_L(\vec R)\Psi_T(\vec R) \ d\vec R$ immediately yields
\be
\int \Psi_T(\vec R)\ Im E_L(\vec R)\Psi_T(\vec R) \ d\vec R=0 \ ,
\lab{gsetup10}
\ee
and, therefore, the only quantity to consider in the following is $Re
E_L(\vec R)$. Before proceeding, we would like to emphasize that the
absolute ground state of the anyon system is a non-analytic function
of $\nu$. Our calculations will simply provide a smooth interpolation.

In the non-interacting case the local energy is
\be
Re E_L(\vec R)=\ds\sum_{k=1}^N[\alpha+\dsf{x_k^2+y_k^2}{2}(1-
{\alpha}^2)+\dsf{\nu^2}{2}(\vec A_{\nu}(\vec
r_k))^2] \ .
\lab{ncnh1}
\ee
The calculation \cite{aormn} of the expectation value for $Re E_L(\vec R)$
gives
\be
E=\frac{N}{2}\left( {\cal N} \ \alpha + \frac{1}{\alpha}\right) \
\lab{ncnh5}
\ee
with
\be
{\cal N}=1+\nu^2 (N-1)[\ln\left(\dsf{1}{2 \delta}\right)- G(N-2)] \ ,
\ee
and $G=3^{1/2}\ln(4/3)$,
which attains a minimum ($\dsf{dE}{d\alpha}=0$) for
\be
\alpha_0={\cal N}^{-1/2} \ .
\lab{ncnh6}
\ee
Thus, the resulting expression for the ground state energy is
\be
E_0=N \ {\cal N}^{1/2} \ .
\lab{ncnh7}
\ee

The logarithmic divergence displayed in $E_0$ when $\delta\rightarrow 0$ has
also been found in other approximate perturbative treatments of the problem
and is widely discussed in the literature (see, for example, Refs. ~\cite{cho,ouvgroup,chitra} ).
Here we have
assumed as in Ref.~\cite{ler} that the cut-off parameter $\delta$ cannot
be zero for $\nu>0$, away from the bosonic limit, since it corresponds
to the square of the nearest distance between the particles (in Ref. \cite{aormn} we have supposed that
$r_0 \approx L$, where $r_0$ is the mean distance between particles). Thus, for
anyons in the parabolic confining potential $\delta$ is definitely
smaller than 1 (in units of $L^2$).

Wu \cite{wu} has computed the ground state energy of $N$ anyons in a 2$D$
harmonic  potential near the bosonic limit $\nu\simeq 0$ and obtained
\be
E \approx [N+N(N-1)\nu/2] \ .
\lab{ncnh8}
\ee
To regularize the expression for $E_0$ we have  made use of this result by
expanding $E_0$, Eq.~\re{ncnh7}, for $\nu \rightarrow 0$ and
identify the leading term in $\nu^2$ with the term linear in $\nu$ of
Eq.~\re{ncnh8}, with the result
\be
\delta= \dsf{1}{2} \exp\left[ -\dsf{1+\nu G(N-2)}{\nu}\right] \ .
\lab{ncnh9}
\ee
With this value of the cut-off parameter the final analytic expression
for the ground state energy is
\be
E_0=N[1+\nu(N-1)]^{1/2} \ .
\lab{ncnh10}
\ee

This formula for large $N$  is
consistent (up to a numerical factor) with the approximate expression
$E \approx \nu ^{1/2}N^{3/2}$ of Chitra and Sen \cite{chitra} calculated
perturbatively from the bosonic end for $\nu>1/N$.

It should be noted, that due to this regularization procedure the ground
state energy obtained, Eq.~\re{ncnh10}, is not an upper bound to the
ground state as one would expect from a variational principle.
This is a consequence of the fitting of
$\delta$ to the result of Wu Ref.~\cite{wu}, which for $N=2$ leads to a
square root dependence in $\nu$, while the exact result for this case
gives a linear dependence. On the other hand, Eq.~\re{ncnh10} applies
for the whole range of parameters of the system $N, \nu$, and
$\omega_0$.

We now include the effect of the Coulomb repulsions between anyons
\be
\dsf{L}{2a_B}\ds\sum_{k,j\not=k}^N\dsf{1}{|\vec r_{kj}|}
\lab{aa1}
\ee
in the
expression for the real part of local energy $Re E_L(\vec R)$, Eq. ~\re{ncnh1}.

The Coulomb interaction part contributes with $N(N-1)$ integrals of the
form $\int \Psi_T(\vec R)\ \dsf{1}{|\vec r_{kj}|}\Psi_T(\vec R) \ d\vec
R$. These integrals have been  evaluated in Ref. \cite{aormn} and
the averaged (real part of the) local energy is
\be
E=\frac{N}{2}\left( {\cal N} \ \alpha + \frac{1}{\alpha}+2{\cal M}
\ \alpha^{1/2}\right)\ ,
\lab{15}
\ee
with
\bea
\ba{l}
{\cal M}=\left(\dsf{\pi}{2}\right)^{1/2}
\dsf{N-1}{2}\dsf{L}{a_B} \ ,
\lab{16}
\ea
\eea
where $a_B$ is the Bohr radius.
The extremum condition $\dsf{dE}{d\alpha}=0$ leads to the equation
\be
X^4-{\cal M}X-{\cal N}=0
\lab{17}
\ee
for $X=1/\alpha^{1/2}$. The
minimum energy is given by the expression
\be
E_0=\dsf{N}{2}\left[\dsf{\cal N}{X_0^2}+X_0^2+\dsf{2{\cal M}}{X_0} \right]
\lab{18}
\ee
and it is achieved at the point
\be
X_0=(A+B)^{1/2}+[-(A+B)+2(A^2-AB+B^2)^{1/2}]^{1/2} \ ,
\lab{19}
\ee
where
\bea
\ba{l}
A=\left[{\cal M}^2/128+\left(({\cal N}/12)^3+({\cal M}^2/128)^2
\right)^{1/2} \right]^{1/3} \ ,\\
B=\left[{\cal M}^2/128-\left(({\cal N}/12)^3+({\cal M}^2/128)^2
\right)^{1/2} \right]^{1/3} \ .
\lab{20}
\ea
\eea
Again, the ground state energy $E_0$, Eq. \re{18}, has a logarithmic
divergence in the limit $\delta \rightarrow 0$ and the quantity  $\cal N$
should be regularized.

In order to determine the cut-off parameter $\delta$, and due to the
lack of analytic results, we had to fit to known numerical results for
the ground state energy at special values of the parameter $L/a_B$.

\begin{figure}
\begin{center}
\includegraphics[angle=0,width=10.5cm,scale=1.0]{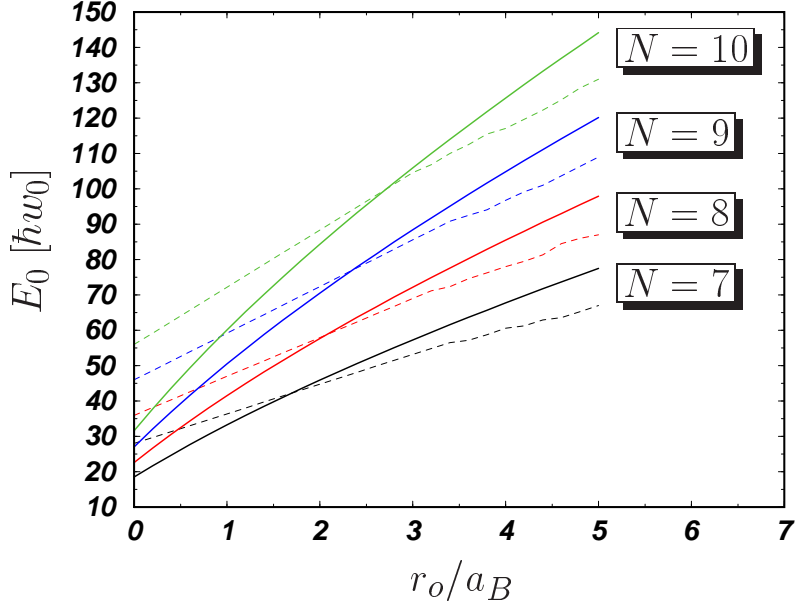}
\end{center}
\caption{
Coulomb interaction parameter $L/a_B$ ($r_0 \approx L$, \, \protect\cite{aormn}) dependence of the ground
state energy for 7 -- 10 electrons calculated by variational
\protect\cite{b1} and  fixed-node quantum Monte Carlo methods
\protect\cite{b3} - dashed curves (results of both calculations are
indistinguishable in these curves) and by formula \protect\re{18} -
solid curves.}
\lab{fig1-2}
\end{figure}

In Figure~1 we have compared the ground state energies calculated for 7-10
electrons using Eq. \re{18}, with the {\it non-interacting}  ${\cal
N}=1+\nu(N-1)$, to variational \cite{b1} and fixed-node quantum Monte
Carlo calculations \cite{b3}.

\section{Harmonic potential regularization}

In the Section 2 we have outlined a variational procedure for the ground
state energy of confined interacting anyons (starting from the bosonic
end) and achieved, after regularization of a logarithmic expression by
means of a cut-off parameter for the particle-particle interaction,
approximate analytic formulas in terms of $N$, $\nu$ and $L/a_B$.  For the
non-interacting anyon system we found
\be
E_0(N, \nu )=\hbar \omega_0 N {\cal N}^{1/2} \ ,
\lab{ggsetup3}
\ee
while for the interacting anyon system the approximate analytic ground
state energy is given by
\be
E_0 (N, \nu )=\dsf{\hbar \omega_0 N}{2}\left[\dsf{\cal N}{X_0^2}+X_0^2+
\dsf{2{\cal M}}{X_0} \right] \
\lab{ggsetup4}
\ee
with the expression for $X_0$, Eq. ~\re{19}.
In these expressions we have used ${\cal N}=1+\nu(N-1)$ and Eq. ~\re{16} for ${\cal M}$.

In order to obtain the corresponding expressions for the 2$D$ homogeneous
anyon gas, which have been derived in the paper \cite{aom} and will be
briefly outlined in this section,  we had to flatten out the parabolic confining
potential while increasing  the number $N$ of anyons, keeping the
density $\rho = N/S=1/\pi r_0^2$ constant, i.e.  we performed the
thermodynamic limit while making the confining potential disappearing.
Here $ \pi r_0^2$ is the area of the jellium disc carrying the positive
countercharge and the mean particle distance  $r_0=a_B r_s$ can be
expressed in units of the Bohr radius by the dimensionless density
parameter $r_s$.

Without Coulomb interaction and in the case of fermions ($\nu=1$) the
ground state energy of the 2$D$ homogeneous electron system of density
$\rho $ is determined by the Pauli exclusion principle and is given
by
\be
E_0(\rho)=\pi \hbar^2 \rho N/M \ ,
\lab{ggsetup8}
\ee
while from \re{ggsetup3} we have
\be
E_0(N, \nu=1 )=\hbar \omega_0 N^{3/2} \ .
\lab{ggsetup9}
\ee
In the thermodynamic limit both expressions have to become identical
and we obtain the relation
\be
\omega_0(N)= \pi \hbar \rho /(M N^{1/2}) \ ,
\lab{ggsetup10}
\ee
which means that, in fact, the thermodynamic limit ($N\rightarrow
\infty$) is obtained for vanishing parabolic confinement potential. We
can extend this consideration for the fermionic limit, $\nu=1$, to the
general anyon case,  $\nu \neq 1$, by assuming instead of \re{ggsetup8}
the relation
\be
E_0(\rho, \nu)=\pi \hbar^2 \rho N \phi(\nu)/M \ ,
\lab{ggsetup11}
\ee
where the function $\phi(\nu)$ is still to be determined under the
constraint  $\phi(\nu=1)=1$. This form is in the accordance with the fact that
close to the bosonic  limit ($\nu \simeq 0$) the ground state energy of
the infinite anyon gas depends  linearly on $\nu$  \cite{sen,wen,mori}.
Consequently, we have for this case
\be
\omega_0(N, \nu)= \pi \hbar \rho f(\nu) /(M N^{1/2})
\lab{ggsetup12}
\ee
with another unknown function $f(\nu)$  and the constraint
$f(\nu=1)=1$.  It turns out $\phi(\nu)$  is determined  by  $f(\nu)$.

In the thermodynamic limit and including the Coulomb interaction, the
parabolic confinement  has to be replaced by the jellium contribution,
which for a disc of radius $R_0$  (containing $N$ counter charges)
gives a potential energy contribution \cite{laupr}
\be
V(\vec r_k)=-\rho \int_{S} \dsf{ e^2 \ d^2 r}{|\vec r_{k}-\vec r|} \ .
\lab{ggsetup13}
\ee
Here $S=\pi R_0^2$ is the area of the jellium disc for $N$ charges and
we have  $R_0=N^{1/2} r_0$. For $\nu = 1$, there is a relation between
characteristic length  $L$ of the oscillator with the mean particle
distance  $L=N^{1/4}r_0$ and we have found $r_0\ll L\ll R_0$
for $N\gg 1$.

In the general case of $\nu \not= 1$ and the Coulomb interaction,
the  approximate analytic expression for the ground
state \re{ggsetup11} can be written in the form
\be
E_0(\nu, r_s)=\pi \hbar^2 \rho N \phi(\nu, r_s)/M \ ,
\lab{ggsetup14}
\ee
thus, in a similar way to generalize $f(\nu)$ by $f(\nu, r_s)$.

Before deriving  the approximate analytic expression for the ground state energy of
Coulomb interacting anyon gas, in Ref. \cite{aom} we have obtained the formula for one of
the 2$D$ Coulomb Bose gas at high densities. Performing the calculation for 2$D$ case, which is
the analog of Foldy's one, Ref. ~ \cite{foldy}, for the 3$D$ case,
we have found
\be
\dsf{E}{N}=- c_{BG}r_s^{-2/3} \ ,
\lab{2dbg8}
\ee
where $c_{BG}=1.29355$. Thus, we had an exact analytic expression for
the ground state energy per particle for the 2$D$ Coulomb Bose gas
valid at high densities. It has next used together
with the known expression for the 2$D$ Coulomb gas in the low density
limit (the 2$D$ Wigner crystal) to derive a form for the unknown
function $f(\nu, r_s)$ introduced above in this section.

In the paper Ref. ~\cite{aom} we have also obtained the spectrum of collective excitations
of the  2$D$ Coulomb Bose gas. It has the form
\be
E_k=[\hbar^2 2\pi e^2\rho k/M+(\hbar^2k^2/(2M))^2]^{1/2} \ ,
\lab{2dbg9}
\ee
which for small $k$ is the 2$D$ plasmon dispersion  and
approaches for large $k$ the free particle dispersion.

For the derivation of an approximate analytic expression  for the ground state energy of
the Coulomb anyon gas,  in paper \cite{aom} we have at beginning calculated the contribution
of the jellium term, Eq. ~ \re{ggsetup13}, into expectation value for energy. We have obtained
\be
\rho \ds\sum_{k=1}^N\int_{S} \int_{-\infty}^{\infty} \Psi_T(\vec r_k)\
\dsf{e^2 \ d^2 r}{|\vec r_k-\vec r|}\Psi_T(\vec r_k) \ d^2 r_k=\dsf{e^2}{L}
N^2 \pi^{1/2} \alpha^{1/2} \ .
\lab{gsqa5}
\ee
It was achieved at the condition  $\frac{\alpha R_0^2}{2L^2} \ll 1$. A validity of this
condition had been shown in \cite{aom} for almost all numerical  values of $\nu$ and $r_s$.

To bring  the contribution of the jellium term in our approximation together
with the Coulomb interaction evaluated in \cite{aormn} and outlined in the previous section
we might write the ground state
energy as in  Eq. \re{ggsetup4}  but with $\hbar \omega_0=\hbar^2/(ML^2)$ and
\bea
\ba{l}
{\cal M}=\dsf{\pi^{1/2}N L}{a_B}\left(\dsf{1}{2^{3/2}}-1 \right) \ .
\lab{gsqa7}
\ea
\eea
However, this expression for ${\cal M}$ we have replaced (the reason for that see \cite{aom}) by
${\cal M}=-c_{WC} N^{1/2} L/a_B$, where the constant $c_{WC}$ has been  fixed by
the ground state energy of the Wigner crystal (limit $r_s \gg 1$).

For sufficiently large $N$, we  used  ${\cal M}=-c_{WC} N^{3/4} r_s$ and
${\cal N}=\nu N$ to write the expression for the ground state energy,
Eq. \re{ggsetup4}, with $X_0=N^{1/4}K_X$, where
\be
K_X=(K_A+K_B)^{1/2}+[-(K_A+K_B)+2(K_A^2-K_A K_B+K_B^2)^{1/2}]^{1/2} \ ,
\lab{gsqa6a}
\ee
and
\bea
\ba{l}
K_A=\left[K^2/128+\left((\nu/12)^3+(K^2/128)^2
\right)^{1/2} \right]^{1/3},\\
K_B=\left[K^2/128-\left((\nu/12)^3+(K^2/128)^2
\right)^{1/2} \right]^{1/3}, \
\lab{gsqa9}
\ea
\eea
with $K=c_{WC} r_s$. It took  the form ( in $Ry$ units)
\be
{\cal E}_0(\nu, r_s)=\dsf{2}{r_s^2}\left[\dsf{\nu}{2K_X^2}+\dsf{K_X^2}{2}-
\dsf{K}{K_X} \right] \ .
\lab{gsqa10}
\ee
If we remember that we have introduced a function
$f(\nu,r_s)$ in $L$, therefore, now $K=c_{WC} r_s/f^{1/2}(\nu,r_s)$ and
the final expression for the ground state energy per particle is
\be
{\cal E}_0(\nu, r_s)=\dsf{2f(\nu,r_s)}{r_s^2}\left[\dsf{\nu}{2K_X^2}+
\dsf{K_X^2}{2}-
\dsf{K}{K_X} \right] \ .
\lab{gsqa11}
\ee
We have determined the function $f(\nu,r_s)$  by fitting
Eq. ~\re{gsqa11} to the known asymptotic limits for the ground state energy
of spin polarized electrons and Coulomb 2$D$ Bose gases  at very small and
very big values of $r_s$. At very big $r_s$ the ground state energy does not
depend on  statistics and equals the energy of the classical 2$D$  Wigner
crystal \cite{bm}, $E_{WC}=-2.2122/r_s$. For the Bose gas at small $r_s$ we
used Eq. ~\re{2dbg8}.

For the bosons at  $\nu=0$  we had from Eq. ~\re{gsqa11}
\be
{\cal E}_0(0, r_s)=-\dsf{c_{WC}^{2/3} f^{2/3}(0,r_s)}{r_s^{4/3}}
\lab{gsqa12}
\ee
with
\be
f(0,r_s)\approx \dsf{c_{BG}^{3/2}r_s/c_{WC}}{1+c_{BG}^{3/2}r_s^{1/2}/c_{WC}} \ ,
\lab{gsqa13}
\ee
where  $c_{WC}^{2/3}=2.2122$  or $c_{WC}=3.2903$.

The asymptotics of the ground state energy, Eq. ~\re{gsqa11}, for $\nu\not=0$
and at big $r_s$  has the form
\be
{\cal E}_0(\nu, r_s \rightarrow \infty ) =\dsf{c_{WC}^{2/3} f^{2/3}(\nu,r_s)}
{r_s^{4/3}}\left(-1+
\dsf{7\nu f^{2/3}(\nu,r_s)}{3c_{WC}^{4/3}r_s^{4/3}}\right) \ .
\lab{gsqa15}
\ee
The function $f(\nu,r_s)$  has to fulfil the constraints:
$f(\nu=1,r_s=0)=1$ for the dense ideal Fermi gas; $f(\nu,r_s=0)=\nu ^{1/2}$ for
the ideal anyon gas close to the bosonic limit (see \cite{sen,wen,mori} ), and $f(0,r_s)$ given by
Eq. \re{gsqa13} for the 2$D$ Coulomb Bose gas. The form
\be
f(\nu,r_s)\approx \nu^{1/2}(1-r_s) e^{-r_s}+ \dsf{c_{BG}^{3/2}r_s/c_{WC}}
{1+c_{BG}^{3/2}r_s^{1/2}/c_{WC}} \ ,
\lab{gsqa17}
\ee
is consistent  with these requirements.

Using this form of  $f(\nu,r_s)$ in Eq. \re{gsqa11}, we have  calculated the ground
state energy per particle in the boson ($\nu =0$) and fermion ($\nu =1$) limits
and compared the results for the latter case with the data obtained by Tanatar
and Ceperley \cite{tanatar} for the spin polarized electron system.

If we look at the results for large $r_s$ in Figure~2 with ground state
energies calculated from Eqs. \re{gsqa11} and  \re{gsqa15}  (the results
coincide on the plotted scale, solid line) and from Eq. \re{gsqa12}
(dash-dotted line) in comparison with the data of \cite{tanatar}, we see the results from
our approximate analytic formula are very close to the "exact" data of
\cite{tanatar}  for the spin-polarized 2$D$ electron system obtained numerically.

\begin{figure}
\begin{center}
\includegraphics[angle=0,width=8.5cm,scale=1.0]{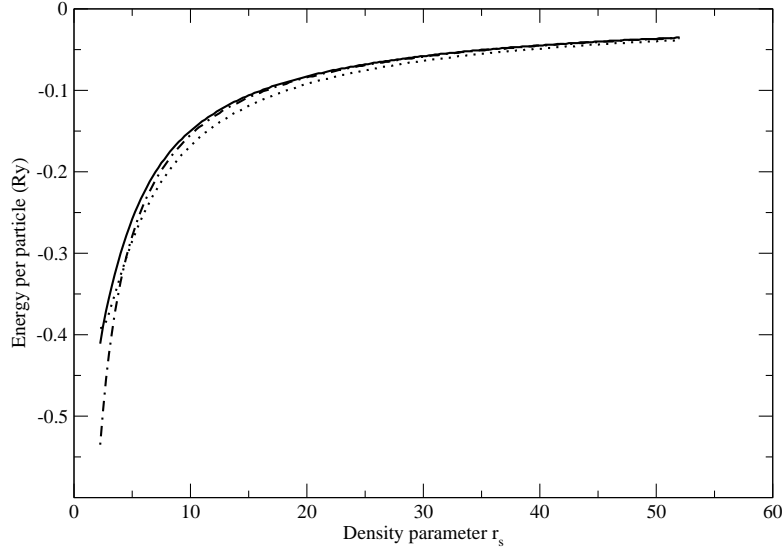}
\end{center}
\caption{
Ground state energy of 2$D$ spin-polarized electrons vs. density parameter
$r_s$ calculated from Eqs. ~\protect\re{gsqa11} or ~\protect\re{gsqa15} for
$\nu=1$  (solid line) in comparison with the data of Ref. ~\protect\cite{tanatar}
(dotted line). The dash-dotted line is obtained from Eq. ~\protect\re{gsqa12}
for the ground state energy of the 2$D$ Coulomb Bose gas.}
\lab{fig2}
\end{figure}

\section{Implicit anyon or single particle boson mechanism of HTCS and pseudogap regime}

We outline in this section results of paper \cite{abdullaev}, where it has been shown that
inside of the ground state the interaction of spins of anyons with statistical magnetic field can induce the
bosonization of these particles, i.e. the transformation of anyons (as also 2$D$ fermions ) into bosons.

In the paper \cite{abdullaev} we have considered the gas of $N$ spinless anyons, which was described by
the Hamiltonian, Eq. \re{gsetup1}, with positive background jellium term, Eq. ~\re{ggsetup13}, included inside
of the first sum.

We have  introduced in the Hamiltonian the term
\be
\dsf{\hbar }{M}\ds\sum_{k=1}^N {\hat {\vec s}} \cdot \vec b_k
\lab{gsetp8}
\ee
with statistical magnetic field \cite{ler,dunne}
\be
\vec b_k = -2\pi \hbar \nu \vec e_z \ds\sum_{j(k\not =j)}
\delta ^{(2)}( \vec r_k- \vec r_j) \ ,
\lab{gsetp9}
\ee
which can be derived if calculates $\vec b_k = \vec \nabla \times
\vec A_{\nu }(\vec r_k)$ by using Eq. ~\re{gsetup2}.
The sign in Eq. ~\re{gsetp8} is taken for electrons with
charge $e=-|e|$. It is chosen to minimize the energy \cite{landau}. For holes
with charge $e=|e|$, we need to change the sign for $\nu$ in Eqs. \re{gsetup2}
and \re{gsetp9}, then Eq. ~\re{gsetp8} and  the expectation value for
energy (see below Eq. ~\re{gsetp12}) retain the sign.

If we take $s_z=\hbar /2$ and take into account that length unit is $L$, so
$\delta ^{(2)}( \vec r )$ should be replaced by $\delta ^{(2)}( \vec r )/L^2$,
then
\be
\dsf{\hbar }{M}\ds\sum_{k=1}^N {\hat {\vec s}} \cdot \vec b_k=
 -\pi  \nu \dsf{\hbar^2}{ML^2} \ds\sum_{k,j(k\not =j)}
\delta ^{(2)}( \vec r_k- \vec r_j) \ .
\lab{gsetp10}
\ee

The calculation of the expectation value, Eq. ~\re{gsetup3}, with the Hamiltonian,
Eq. ~\re{gsetp10}, and wave function $\Psi_T(\vec R)$,  Eq. ~\re{gsetup7},
gives (in $\hbar^2/(ML^2)$ units)
\be
-\pi  \nu \ds\sum_{k,j(k\not =j)} \int \Psi_T(\vec R)\
\delta ^{(2)}( \vec r_k- \vec r_j) \ \Psi_T(\vec R) \ d\vec R =
\dsf{-\nu \alpha N(N-1)}{2} \ .
\lab{gsetp11}
\ee

The total expectation value for the ground state energy  including all terms
of the Hamiltonian  is
\be
E= \frac{N {\cal N} \ \alpha }{2} + \frac{N}{2\alpha}+
N{\cal M}
\ \alpha^{1/2} - \dsf{\nu \alpha N(N-1)}{2} \ ,
\lab{gsetp12}
\ee
where the quantity ${\cal M}$  contains the Coulomb interaction and
jellium background term (see previous section).

As in previous sections we have used the expression ${\cal N}=1+\nu(N-1)$ for
${\cal N}$, which is result of the cut-off parameter regularization \cite{aormn}
of the logarithmic divergence,  revealing in the bosonic representation, when two
particles come up to each other.  Substituting this expression for
${\cal N}$ into  expectation value for energy $E$, Eq. ~\re{gsetp12}, we
see the exact cancellation of terms containing the $\nu$ factor. This can be also
achieved  putting in the formula for ground state energy $\nu=0$, i.e.
the case of bosons. As the energy of bosons is lower than one for fermions or anyons
with $\nu \not= 0$, there appears a coupling of spin with statistical magnetic field for every
particle or bosonization of 2$D$ fermions and anyons. Of course, this effect occurs
when the particles have the spin $\hbar /2$.

As next step of this coupling scenario might be the fluctuations of spins of bosons around of
statistical magnetic field. Hence, the bosons might reveal an effective spin and
look like as Fermi particles. However, Fermi gases with different spins are
independent \cite{landaus}. Thus, we have presumed that in the zero order
approximation the spins of bosons interact only with each other and do not
interact with spins of another fermions if they exist in the system.
Interaction of spins might take place inside of some length- spin correlation
radius, which for temperature $T=0$ we denote as $\xi_0$. The final feature
of this model might be a destruction of spin and magnetic field coupling at
the increasing of spin fluctuations when bosons become the fermions. This
would occur when the gain in the energy of bosons due to fluctuations is equal
to energy difference between the Fermi and Bose ground states.

The interaction term of boson spins in the Hamiltonian we brought in the form
\be
e^{-r_0/\xi_0 }  \ds\sum_{k=1}^N {\hat {\vec s}}_{k+\delta} \cdot
{\hat {\vec s_k}} \ .
\lab{gsetp13}
\ee
Here $r_0$ is the mean distance between particles. We introduced in this expression
a factor $e^{-r_0/\xi_0 }$, which takes into account the exchange character of
interaction of spins \cite{landau1}. However, if the typical scale of  $\xi_0$
is the nearest interatomic  distance, for our sreened by magnetic field spins,
$\xi_0$ is to be assumed phenomenological and taken from experiment.

For the $T=0$ case, a following argument allowed us to establish the explicit
form of Eq. ~\re{gsetp13}. The boson ground state energy was obtained when
the term, Eq. \re{gsetp8}, was included in the Hamiltonian. To get again the
fermion ground state energy, we need to cancel it. Therefore, for dense
($r_0<\xi_0$) boson gas, there should be
${\hat {\vec s}}_{k+\delta}=-\hbar \vec b_k/M$. Comparing with
Eq. ~\re{gsetp9} we see that the
summation index $\delta$ includes all $N-1$ numbers of index $j$.

The expression for the Hamiltonian of bosonized infinite
Coulomb anyon gas with interaction of spins is
\be
\hat H=\dsf{1}{2M}\ds\sum_{k=1}^N\left[\left(\vec p_k+\vec
A_{\nu}(\vec r_k)\right)^2+M V(\vec r_k)\right] +\dsf{1}{2}\ds\sum_{k,j\not=k}^N\dsf{e^2}
{|\vec r_{kj}|}+
\dsf{\hbar (1-e^{-r_o/\xi_o }) }{M}\ds\sum_{k=1}^N {\hat {\vec s}} \cdot
\vec b_k  \ .
\lab{gsetp14}
\ee

We can bring here the results of Ref. \cite{aom} and previous section.
The ground state energy of the Coulomb anyon gas at $r_s>2$ with the high accuracy (see Figure~2)
is described by formula, Eq. ~\re{gsqa15}, with the expression for function
$f(\nu,r_s)$,  Eq. ~\re{gsqa17}. For the HTCS, actual for us is the region $r_s>9$ (see below Figure 4 ),
where one can put $f(\nu,r_s)\approx f(\nu=0,r_s)$. The ground state energy for the 2$D$ Bose and Fermi
gases is obtained putting in  Eq. ~\re{gsqa15} $\nu=0$ and $\nu=1$, respectively.

The analogical to Ref.  ~\cite{aom} and Section 3 calculation, with the Hamiltonian, Eq. ~\re{gsetp14},
gave for $r_s>2$ the same expression, Eq. ~\re{gsqa15}, for
ground state energy of bosonized anyons, but in it one needs to replace $\nu$ by
$\nu e^{-r_s/\xi_0 }$  (now $\xi_o$ is expressed in $a_B$ units ).
Considering below the bosonization of 2$D$ fermions  we have putted now $\nu=1$.

To become the pure fermions bosons had to overcome the energy difference
\be
\Delta_0^B = \dsf{7 (1- e^{-r_s/\xi_0 }) f^{4/3}(0,r_s)}
{3c_{WC}^{2/3}r_s^{8/3}} \ .
\lab{gsetp17}
\ee

One needs  a special remark here. Our previous calculations have been related
to ground state of spinless or fully spin polarized fermions (electrons).
It would be preferable to deal with one of normal, i.e.
with no spin polarized, electron liquid state. However, the difference of their
ground state energy is essentially lower \cite{tanatar} than accuracy of our
calculations.

We have applied the model for possible clarification of summarized experimental phase
diagram of hole doped High-$T_c$ superconductors, which was recently
proposed by Tallon and Loram \cite{tallon}. As it was said above it is believed that the main
contribution into HTCS is provided by $ab$ planes
of set of $CuO_2$ atoms and therefore, 2$D$ is responsible for it. The SC
state occurs when holes transit into BEC. We note that for the ideal 2$D$
systems BEC can exist only at $T=0$ \cite{lifshic}. At $T\not=0$ the evolution
of fluctuations for the order parameter destroys BEC. However, if 2$D$ gas
is situated inside of third dimension this evolution  is suppressed (with
respect to SC see book of Abrikosov \cite{abrikosov}). Furthermore, we used
in the Hamiltonian, Eq. ~\re{gsetp14}, a three dimensional Coulomb potential.
Therefore, for HTCS the 2$D$ BEC of bosons can also exist at not
zero temperature.

We used $\Delta_0^B$ as gap for SC and studied its asymptoticses as function of
density of holes $n_s=1/(\pi r_s^2)$. Below we show that $n_s\sim p$,
where $p$, called in the experiment as concentration of holes, is the
fractional part of hole per atom $Cu$. At  big values of $r_s$ or small
$n_s$ one can neglect the exponential factor in Eq. \re{gsetp17} and
$\Delta_0^B\sim n_s\sim p$. At  small $r_s$ or big $n_s$, without this
factor, $\Delta_0^B$ would have $\Delta_0^B\sim n_s^{2/3}\sim p^{2/3}$, but
$e^{-r_s/\xi_0 }$ suppresses this dependence to zero and the law of it
depends from function  $\xi_0(p)$. At this limit of $r_s$ we presumed that
the function $\Delta_0^B(p)$ coincides with experimental dependence $E_g(p)$.
Extrapolating this asymptotics of $\Delta_0^B(p)$ to small values of $p$ and
equating it to $E_g(p)$ one finds the empirical dependence $\xi_0(p)$. For it
$\xi_0(p)\sim 1/E_g(p)$.

The experimental values for energy are expressed in the Kelvin temperature, $K$,
units. To be sure in the correctness of our 2$D$ density of holes, we expressed
it in $cm^{-2}$ units and compared with experiment. For elementary structural
cell of almost all cuprate $ab$ planes $a=3.81 \AA$ and $b=3.89 \AA$.
Assuming that it has one atom of $Cu$ with $p$ fraction of hole, the density
is $n_{ab}=N_{ab}\cdot p \  cm^{-2}$, where $N_{ab}$ is number of elementary
cells per $ 1 cm^2$ square. We have
$n_{ab}= 6.7472\cdot  p\cdot  10^{14} \ cm^{-2}$, from where $n_s\sim p$. At
the optimal doping of holes, $p\approx 0.16$, we compared $n_{ab}$ with
experimental one of \cite{puchkov}. In it for Y-123 compound the square of three
dimensional plasma frequency, being expressed in square of wave number units,
for SC carriers has a value $1.1\cdot 10^8 \ cm^{-2}$. We find the density of
holes per $ 1 \ cm^3$ and then, assuming that for Y-123 compound there are two
$CuO_2$ planes \cite{dagotto} per elementary cell, responsible for SC, and
in the $c$ axis the spacing is $c=11.7 \AA$, transform it into 2$D$
$n_{ab}^{exp}= 0.9137\dot 10^{14} \ cm^{-2}$. Our optimal doping value for
$n_{ab}$ is $1.0795\cdot 10^{14} \ cm^{-2}$. Therefore, typical density for
SC carriers for the optimal doping state is $\sim 10^{14} \ cm^{-2}$. From the
experiment \cite{puchkov} also leads important information about approximate
equality of SC carriers density to whole one. Thus, we can bring $n_{ab}$ as
SC density. A comparison of this value for $n_{ab}$  with
$n_{FQHE}\sim 10^{11} \div 10^{12}\ cm^{-2}$ for Fractional Quantum Hall Effect
(FQHE) experiment \cite{chang} shows importance of the pure, not screened,
Coulomb potential interaction, which is essence of FQHE \cite{laupr},
for HTCS. The screened Coulomb potential is supposed to be as justification
\cite{anderson} for the treatments based on the Hubbard model.

On the Figure~3 we presented the summarized curve for pseudogap boundary energy
$E_g$ (Fig. 11 from paper \cite{tallon}), as result of interpolation of sets of
experimental data,  SC gap energy $\Delta_0=4 K_B \ T_c$, which was evaluated
by empirical formula $T_c=T_{c,max}[1-82.6(p-0.16)^2]$
with $T_{c,max}=95 \ K$ for $Bi$ \ 2212 compound \cite{presland}, and
calculated by formula  Eq. ~\re{gsetp17}  SC gap energy for bosons
$\Delta_0^B$ as function of concentration of holes $p$. Despite a
$\Delta_0^B(p)$ dependence has the qualitatively different form  than
experimental one, it is in accordance with the conclusion of Tallon and Loram
that $E_g$, being going to zero at the critical doping concentrate
$p_c\approx 0.19$, separates BEC into two parts, where in first one the
density of SC providing pairs is small (even though that conclusion is not expressed explicitly
by Tallon and Loram, however, Fig. 10 of their paper more than obviously displays it). As we see
from the figure the magnitudes of this and experimental diagrams have the same order. We would like
to note that, apparently, it is the first estimate for BEC energy of canonical
bosons, which can be appropriate for experiments of SC.

Figure~4 demonstrates the $p$ dependencies of $r_s$ and $\xi_0$. The spin
correlation radius $\xi_0$ becomes to be sharply increased when $p$
approaches $p_c$, which might mean the vicinity of phase transition, where
many and many spins of bosons are involved in correlation process before
the transforming of them into fermions. We established a correspondence with
second important conclusion of Tallon and Loram that "the pseudogap is
intimately connected with (though not equivalent to) short-range Anti
Ferromagnetic (AF) correlations, which disappear at the same critical doping
state" ($p_c$), and results shown in Figure~4.
The paper \cite{tallon} shows that experimental short-range AF correlations
scale like $E_g(p)$ dependence and vanish at $p_c$ (see Fig. 6 of this
paper). It was also suggested in Ref. \cite{tallon} that two these quantities
are the temperature independent. Our treatment is not for the temperature
$T\not=0$. However, last remark allow us to do the connection between
short-range AF correlations and correlation radius $\xi_0$.
The presumption about correlations of spins of bosons would mean that
there would exist the competing of these correlations  with AF correlations,
where nature of last ones is the spin interaction of closely situated fermions.
Increasing of correlation radius $\xi_0$ would lead to revealing the Fermi
like spin correlations of bosons in the extended region of sample and thus,
assuming that in the first order approximation these spins now interact with
ones of fermions, to suppressing of short-range AF correlations inside of
this region. From this a disappearing of AF correlations at $p_c$, where
$\xi_0$ goes to infinity, would be natural and might be considered as indirect
indication of destruction of BEC from canonical bosons. Mathematically the
short-range AF correlations will be proportional to $1/\xi_0(p)$, because
$E_g(p)\sim 1/\xi_0(p)$. One might suppose that the interaction of spins of
fermions with statistical magnetic field (as well as the statistical magnetic
field itself) of anyons would have a hidden character and might not be
revealed  obviously in the experiment. In this case, the described
experimental behaviour of short-range AF correlations might be also
the implicit indication of existing of the statistical magnetic field.

Next conclusion of Tallon and Loram paper \cite{tallon} is an independence of
pseudogap boundary energy $E_g$ and phase diagram for BEC. As we said above,
in our treatment, without $E_g$, $\Delta_0^B$ would have a proportionality
$\Delta_0^B\sim p^{2/3}$. The phenomenological including of $E_g$ determines
the maximum $T_c$ value for BEC phase and critical concentration $p_c$. Close
to $p_c$ $E_g(p)$ merges with $\Delta_0^B(p)$ determining the asymptotics for
latter one. In our model $E_g$ characterizes the cuprate material. As for
almost all High-$T_c$ superconducting samples the numerical values for
cell spacing constants $a, b$ are the same \cite{dagotto}, according to our
approach we should been obtain the similar $T_{c,max}$ for all of them. The
value for $T_{c,max}$ depends also from number of $CuO_2$ planes in the
elementary cell \cite{dagotto}. We have considered the sample with one plane
in the hoping that $E_g(p)$ takes effectively into account this factor.

Here we would like to say some words about the  possible scenario for the  phase diagram
of HTCS, which might be  derived from this model. Due to vicinity of the
structural phase transition  to superconducting state in the cuprate materials at the
concentration of holes below optimal \cite{dagotto}, we might do a hypothesis
that the induced by it the mechanical strain  would strengthen a quadratic
striction and therefore, the phase transition into BEC would be not a second order, as it
should be, but first, close to second one \cite{landau2}. For this case a
pseudogap regime would find natural interpretation. It would correspond to
meta stable phase of bosons. At pseudogap boundary, $E_g$, bosons would finally
undergo a phase transition into fermions. The critical concentration, $p_c$, at which the
pseudogap energy, $E_g$, is zero, would be considered as critical point of
first order phase transition. The first order phase transition scenario is possibly consistent with recent
experimental observations of coexistence of pseudogap and superconducting
phases described in the review of Pines \cite{pines}, because typically the
coexistence of equilibrium phases exists in this order of phase transitions \cite{landau3}.
In our case, these phases would be the small islands of meta stable phase of
bosons and BEC situated in close area to BEC boundary in the phase diagram.
For electron doped materials one may suppose that there is no structural phase
transition. Therefore, the phase transition into BEC would have a pure second order,
which is possibly seen in the experiments. At last, the hypothesis might
clarify the possible independence of pseduogap energy, $E_g$, and BEC suggested
in the paper \cite{tallon}. As it was said above the boundary, $E_g$, would be connected
with mechanical strain property of cuprate materials, while condensation
into BEC with coupling of spins of fermions with statistical magnetic field.
However, the spin correlations of bosons, which are to be governed by $E_g$
and suppress the BEC, might implicitly be relate two these quantities.

\begin{figure}
\begin{center}
\includegraphics[angle=0,width=9.5cm,scale=1.0]{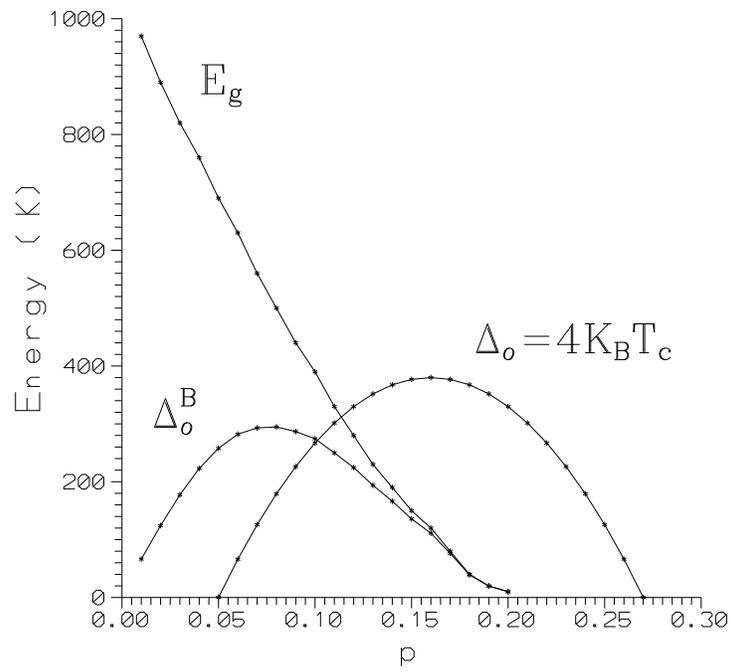}
\end{center}
\caption{
The experimental pseudogap energy $E_g$, SC gap energy $\Delta_0=4 K_B \ T_c$
(experiment for $Bi \ $ 2212 compound), and calculated by formula
Eq. ~\protect \re{gsetp17} SC gap energy for bosons
$\Delta_0^B$ in Kelvin temperature (K) units as function
of concentration of holes $p$.
}
\lab{fig3}
\end{figure}
\begin{figure}
\begin{center}
\includegraphics[angle=0,width=9.5cm,scale=1.0]{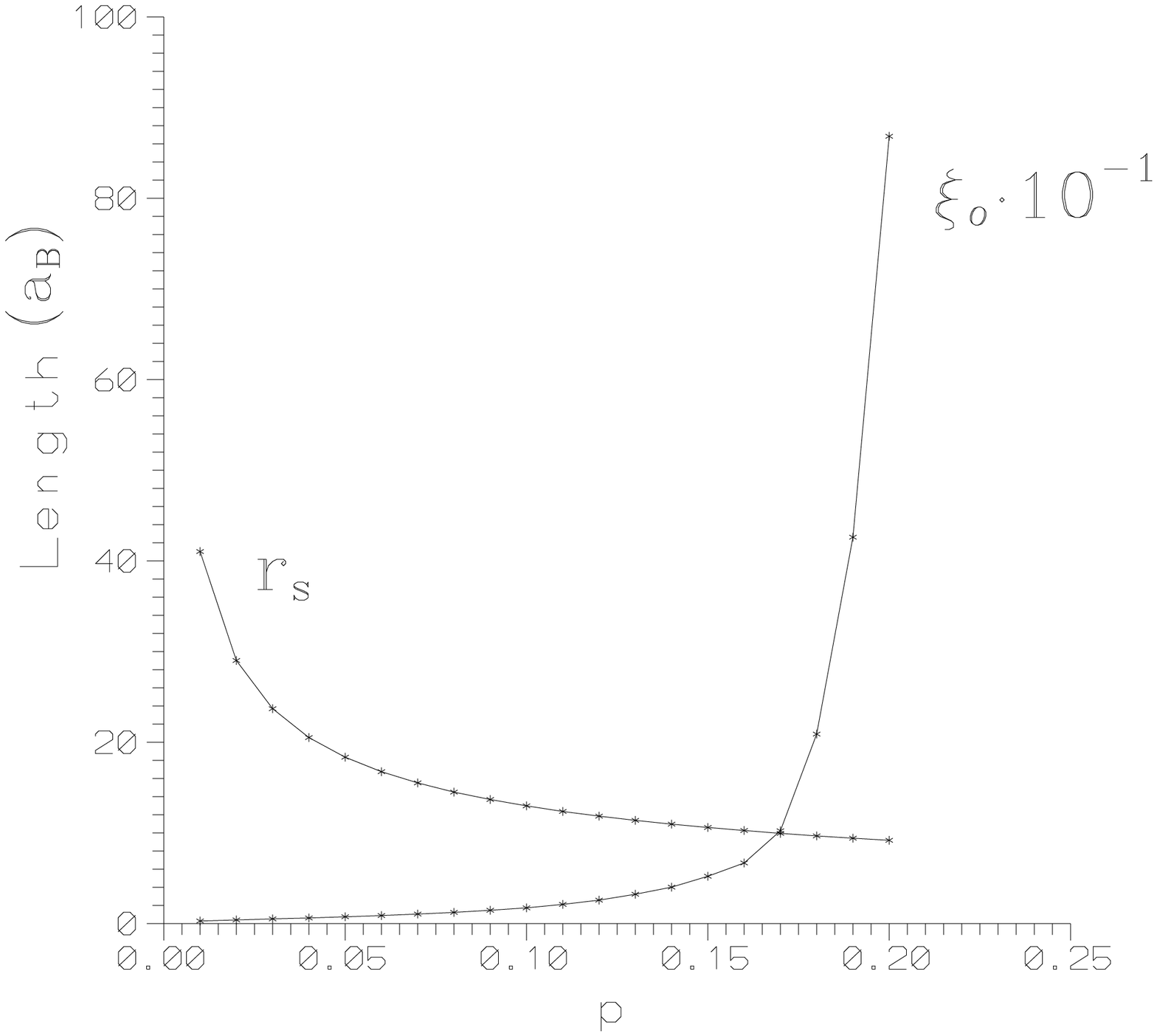}
\end{center}
\caption{
The mean distance between holes $r_s$ and spin correlation
radius $\xi_0$ in Bohr radius $a_B$ units as function of concentration of
holes $p$.
}
\lab{fi4}
\end{figure}

Now about the spectrum of collective excitations, Eq. ~\re{2dbg9}. This spectrum
has no gap at small wave vectors $k$ and
therefore, can not be responsible for SC. SC is provided by particles in the
BEC. On the other hand, one might expect that the bosons with effective  spins
inside of correlation radius $\xi_0(p)$ and obeying the relation,
Eq. ~\re{2dbg9}, at big wave vectors $k$  are not the Fermi liquid
quasiparticles like.

We considered the effect of bosonization of 2$D$ fermions for the instance
of HTCS. One might presume that it would be important for any
2$D$ fermion gas. FQHE, gas of isotopes $He \ $  3 and many other objects
might be affected by this effect.

\section{Summary}

We have introduced in the Hamiltonian of anyon gas the Zeeman term of the
interaction of spins $s_z=\hbar /2$ of particles with magnetic field induced
by anyon vector potential, i.e. statistical magnetic field. A calculation of
the expectation value for ground state energy in the framework of variational
approach with cut-off parameter regularization has exhibited the cancellation
of terms connected with fractional statistics, which might be mean the
bosonization of anyons due to a coupling of their spins with statistical
magnetic field. For fermions, as the particular case of anyons, we have applied
it for the possible clarification of summarized phase diagram of HTCS
suggested by Tallon and Loram for the region below preudogap $E_g$.
Additionally to Zeeman term we have phenomenologically introduced in
Hamiltonian  the term, which was responsible for the correlations of
effective  spins of bosons, and connected them with $E_g$. The obtained phase
diagram has been quantitatively  close to experimental one, while model
qualitatively described conclusions made in paper of Tallon and Loram.

\section{Acknowledges}

Author acknowledges the Volkswagen Foundation for the support of  Section 3,
Korea Research Foundation Grant (KRF-2004-005-C00044) of Section 4,
U. Roessler for the discussions  and critical reading of Sections 2 -- 3,
G. Ortiz for the discussions of Sections 2 -- 3 and C.-H. Park for the discussions
of all sections of the paper.


\begin{thebibliography}{99}

\bibitem{ogg}
R. A. Ogg, \emph{Phys.\ Rev.} \ {\bf 69}, 243 (1946).

\bibitem{shaf}
M. R. Schafroth, \emph{Phys.\ Rev.} \ {\bf 96}, 1149, 1442 (1954);
{\bf 100}, 463 (1955).

\bibitem{shafbb}
M. R. Schafroth, S. T. Burler, and  J. M. Blatt, \emph{Helv.\ Phys.\
Acta} \ {\bf 30}, 93 (1957).

\bibitem{ginzburg}
V. L. Ginzburg, \emph{Phys.\ Uspekhi} \ {\bf 43}, 573 (2000).

\bibitem{dagotto}
E. Dagotto, \emph{Rev. Mod. Phys.} {\bf 66}, 763 (1994).

\bibitem{ler}
A. Lerda, {\it Anyons}\ (Springer-Verlag, Berlin, 1992).

\bibitem{wil1}
F. Wilczek, {\it Fractional Statistics and Anyon Superconductivity}
(World Scientific, Singapore, 1990).

\bibitem{lei}
J. M. Leinaas and J. Myrheim, \emph{Nuovo Cimento B} {\bf 37}, 1
(1977).

\bibitem{wil2}
F. Wilczek,  \emph{Phys. Rev. Lett.} {\bf 48}, 1144 (1982).

\bibitem{ortiz}
There is another opinion (G. Ortiz) that due to the difference of Hilbert spaces for these two
particles there is a strict (mathematical) restriction for this like transformation. The opposite
argument is the approximate ground state calculation of anyons in the harmonic potential \cite{chitra}.
Starting from boson end it reproduces with the high accuracy the energy of fermions. The last argument
supports also the result of Section 3 (see below).

\bibitem{berestecky}
V. B. Berestecky, E. M. Lifshitz, and L. P. Pitaevsky {\it  Quantum
Electrodynamics.} (Nauka, Moscow, 1980) $\S$ 25 (in russian).

\bibitem{abdullaev}
B. Abdullaev, cond-mat/0404668.

\bibitem{dunne}
G. Dunne, A. Lerda, S. Sciuto, and C. A. Trugenberger,  \emph{Nucl.
Phys. B} {\bf 370}, 601 (1992).

\bibitem{tallon}
J. L. Tallon, and J. W. Loram, \emph{Physica C} {\bf 349}, 53
(2001); cond-mat/0005063.

\bibitem{aormn}
B. Abdullaev, G. Ortiz, U. Roessler, M. Musakhanov, and A. Nakamura,
\emph{Phys. Rev. B} {\bf 68}, 165105 (2003).


\bibitem{aom}
B. Abdullaev, G. Ortiz, U. Roessler, and M. Musakhanov, \emph{Phys. Rev. B}, will
be submitted.

\bibitem{wu}
Y. -S. Wu,  \emph{Phys. Rev. Let.} {\bf 53}, 111 (1984),
\emph{Erratum ibid} {\bf 53}, 1028 (1984).

\bibitem{lau}
R. B. Laughlin,  \emph{Phys. Rev. Lett.} {\bf 60}, 2677 (1988).

\bibitem{ceperley}
D. M. Ceperley and M. H. Kalos in {\it Monte Carlo Methods in
Statistical Physics}, edited by K. Binder (Springer-Verlag, Berlin,
1979).

\bibitem{cho}
C. Chou, \emph{Phys. Rev. D} {\bf 44}, 2533 (1991); \emph{D} {\bf
45}, 1433 (1992) (E); C. Chou, L. Hua and G. Amelino-Camelia,
\emph{Phys. Lett. B} {\bf 286}, 329 (1992); G. Amelino-Camelia,
\emph{Phys. Lett. B} {\bf 299}, 83 (1992).

\bibitem{ouvgroup}
A. Comtet, J. McCabe and S. Ouvry, \emph{Phys. Lett. B} {\bf 260},
372 (1991); J. McCabe and S. Ouvry, \emph{Phys. Lett. B} {\bf 260},
113 (1991); A. Dasnieres de Veigy and S. Ouvry, \emph{Phys. Lett. B}
{\bf 291}, 130 (1992); \emph{Nucl. Phys. B} {\bf 388}, 715 (1992).

\bibitem{chitra}
R. Chitra and  D. Sen,  \emph{Phys. Rev. B} {\bf 46}, 10923 (1992).

\bibitem{b1}
F. Bolton, \emph{Phys. Rev. Lett.} {\bf 73}, 158 (1994).

\bibitem{b3}
F. Bolton,  \emph{Solid  State  Electron.} {\bf 37}, 1159 (1994).

\bibitem{sen}
D. Sen and R. Chitra, \emph{Phys. Rev. B} {\bf 45}, 881 (1992).

\bibitem{wen}
X. G. Wen and A. Zee, \emph{Phys. Rev. B} {\bf 41}, 240 (1990).

\bibitem{mori}
H. Mori, \emph{Phys. Rev. B} {\bf 42}, 184 (1990).

\bibitem{laupr}
R. B. Laughlin in {\it The Quantum Hall Effect}, edited by R. E. Prange and
S. M. Girvin (Springer-Verlag, 1987).

\bibitem{foldy}
L. L. Foldy, Phys. Rev. {\bf 124}, 649 (1961); see also in N. M. March,
W. H. Young and S. Sampanthar {\it The many body problems in quantum mechanics}
(Univ. Press, Cambridge, 1967).

\bibitem{bm}
L. Bonsal, and A. A.  Maradudin, \emph{Phys. Rev. B} {\bf 15}, 1959
(1977).

\bibitem{tanatar}
B. Tanatar, and D. M. Ceperley, \emph{Phys. Rev. B} {\bf 39}, 5005
(1989).

\bibitem{landau}
L. D. Landau, and E. M. Lifshitz {\it  Quantum Mechanics. Nonrelativistic
Theory.} (Nauka, Moscow, 1989) $\S$ 112 (in russian).

\bibitem{landaus}
L. D. Landau, and E. M. Lifshitz {\it  Quantum Mechanics. Nonrelativistic
Theory.} (Nauka, Moscow, 1989) $\S$ 65 (in russian).

\bibitem{landau1}
L. D. Landau, and E. M. Lifshitz {\it  Quantum Mechanics. Nonrelativistic
Theory.} (Nauka, Moscow, 1989) task 1 to $\S$ 62 (in russian).


\bibitem{lifshic}
E. M. Lifshitz, and  L. P. Pitaevsky {\it  Statistical Physics.
Theory of Condensed State.} (Nauka, Moscow, 1978) $\S$ 27 (in russian).

\bibitem{abrikosov}
A. A. Abrikosov {\it  The Basic Theory of Metals.} (Nauka, Moscow, 1987)
$\S$ 16.11 (in russian).

\bibitem{puchkov}
A. V. Puchkov, D. N. Basov, and T. Timusk, \emph{J. Phys. Cond.
Mat.} {\bf 8}, 10049 (1996); cond-mat/9611083.


\bibitem{chang}
A. M. Chang in {\it The Quantum Hall Effect}, edited by R. E. Prange and
S. M. Girvin (Springer-Verlag, 1987).

\bibitem{anderson}
P. W. Anderson, P.A. Lee, M. Randeria, T. M. Rice, N. Trivedi, and
F. C. Zhang, cond-mat/0311467.

\bibitem{presland}
M. R. Presland, J. R. Tallon, R. G. Buckley, R. S. Liu, and N. E.
Flower, \emph{Physica  C} {\bf 176}, 95 (1991).

\bibitem{landau2}
L. D. Landau, and E. M. Lifshitz {\it  Statistical Physics.}
(Nauka, Moscow, 1976) $\S$ 146 (in russian).

\bibitem{pines}
D. Pines, cond-mat/0404151.

\bibitem{landau3}
L. D. Landau, and E. M. Lifshitz {\it  Statistical Physics.}
(Nauka, Moscow, 1976) $\S$ 81 (in russian).




\end{thebibliography}
\end{document}